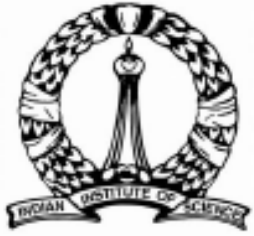 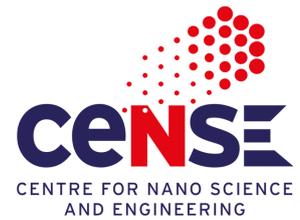

# Fabrication and Characterization of AlN-based, CMOS compatible Piezo-MEMS Devices

By

**Shubham Jadhav**

Centre for Nano Science and Engineering

Indian Institute of Science

Bangalore-560012

JULY 2019

Signature of the Author: . . . . . . . . . . . . . . . . . . . . . . . . . . . . . . . .

**Shubham Jadhav**

**Centre for Nano Science and Engineering**

**Indian Institute of Science, Bangalore**

Signature of the Supervisor: . . . . . . . . . . . . . . . . . . . . . . . . . . . . . . . .

**Prof. Rudra Pratap**

**Professor**

**Centre for Nano Science and Engineering**

**Indian Institute of Science, Bangalore**

# Table of Contents

# CHAPTER 1
# INTRODUCTION

Microelectromechanical systems (MEMS) are miniaturized devices with interactive capabilities that when coupled with IC (integrated circuit) components have the ability to interact with the environment or other on-chip components, allowing the creation of sensor, actuator, and/or transducer systems thereby. There are different types of actuation/Sensing techniques can be implemented for MEMS actuator and Sensors: electrostatic, piezoelectric, thermal and magnetic. But for particular applications like Micromachined Ultrasonic Transducers(PMUTs) and gyroscope; Piezoelectric technique stands out in term of low power consumption, fast switching and robust operation along with simple electronics as compared to other technique. So, in this work, we decided to choose Piezoelectric method for actuation and sensing.

PiezoMEMS are MEMS with piezo as an active component. The piezoMEMS sensor or actuator are widely used in electromechanical, optoelectronic or micro-fluidic systems. Examples are piezoMEMS acceleration sensors, micromirrors, lab-on-a-chip, and Piezoelectric Micromachined Ultrasonic Transducers or PMUTs.

The piezoMEMS microactuators and microsensors are fabricated on a single substrate through the use of thin-film piezoelectric and structural materials. Such thin-film materials are deposited using standard semiconductor thin film techniques such as spin coating, sputtering, and evaporation. The films are patterned and etched using standard photolithography tools and chemical etching methods in order to form the desired device structure.

For these devices to work efficiently, we need good control over the quality of piezoelectric material. Quality of piezo material can primarily be defined by Piezoelectric coefficient, Piezoelectric Coupling Factor and quality factor.

## 1.1 Piezoelectric coefficient

Piezoelectric materials develop charge on the sample surfaces when exposed to applied stresses (direct effect) or equally, under applied electric field, they undergo a change in their dimensions (converse effect). The electric displacement and stress/strain are related by a linear dependence at low fields, expressed by the following equations:

$$D_i = \sum_k d_{ik}\sigma_k \text{ and } D_i = \sum_k e_{ik}x_k \text{(direct effect)},$$

$$x_i = \sum_k d_{ki}E_k \text{ and } \sigma_i = -\sum_k e_{ki}E_k \text{(converse effect)}$$

$D_i$ is the electric displacement and $E_k$ is the electric field tensor; $\sigma_k$ and $x_k$ are the stress and strain tensors, respectively, and $d_{ik}$ and $e_{ik}$ are the piezoelectric coefficient tensors.

## 1.2 Piezoelectric Coupling Factor

The piezoelectric coupling factor ($K^2$) is a unitless quantity that could be used to compare the efficiency of different piezoelectric material in mutually converting electrical energy to mechanical energy. One way to define the piezoelectric coupling factor is as follows (in a loss-free scenario):

$$K^2 = \frac{W_M}{W_M + W_E}$$

Where $W_M$ is the work that is delivered to a mechanical load by a piezoelectric actuator that is preloaded with potential energy ($W_M + W_E$) by connecting it to an electrical source, while it is mechanically free to move. It could be shown that this ratio is a function of material properties as follows (given the stress is only non-zero in one direction or in other words in a 1D scenario):

$$K^2 = \frac{e^2}{c^E \epsilon^T},$$

Where e is the piezoelectric constant (with the unit of C/m²) in the direction of the electric field and mechanical stress, $c^E$ is the stiffness (i.e., elastic constant with the unit of N/m²) at zero electric field, and $e^T$ (with the unit of $\frac{F}{m} = \frac{C}{V \times m}$) is the dielectric constant in the direction of an electric field at zero stress.

The piezoelectric coefficient ($e_{33}$) of AlN is 1.55 C/m². This is about ten times less than that of PZT thin films. However, combined with its roughly 100 times lower dielectric constant, AlN shows a relatively large coupling coefficient for the thickness vibration mode: $k_t^2$=6.5%.

| Material | Quartz | AlN | ZnO | PZT |
|---|---|---|---|---|
| $K^2$ (%) | 0.86 | 6.5 | 8.5 | 23 |

Furthermore, it is a hard material with light atoms and exhibits a high mechanical quality up to very large microwave frequencies[1,2].

The most common material used in PiezoMEMS devices are Pb(Zr$_{1-x}$Ti$_x$)O$_3$, AlN, ZnO, and PVDF.

## 1.3 Material selection

Among the various piezoelectric materials available, each one has its own benefits and limitations. Thus, choosing a material that fulfils all the requirements needed for integrating into a MEMS device is crucial.

As compared to widely used piezoelectric materials PZT and ZnO, AlN fulfils our device requirements better. ZnO is ruled out because it has a high dielectric loss tangent and lower bandgap with a lower breakdown voltage and PZT has a high piezoelectric coefficient but its unstable nature and need for poling make it unsuitable for the device in consideration [3–5].

**Table 1.1:** The following table illustrates the comparison of various physical properties of the various commonly used piezoelectric materials

| | AlN | ZnO | PZT |
|---|---|---|---|
| Density $(kg/m^3)$ | 3230 | 5610 | 7570 |
| Piezoelectric constant $(C/m^2)$ | $e_{31} = -0.58$ | $e_{31} = -0.57$ | $e_{31} = -6.5$ |
| | $e_{33} = 1.55$ | $e_{33} = 1.32$ | $e_{33} = 23.3$ |
| Thermal conductivity $(W/cm \cdot °C)$ | 2.85 | 0.6 | 0.018 |
| Thermal expansion $(1/°C)$ | $\alpha_a = 4.2 \times 10^{-6}$ | $\alpha_a = 6.5 \times 10^{-6}$ | $\alpha = 2 \times 10^{-6}$ |
| | $\alpha_c = 5.3 \times 10^{-6}$ | $\alpha_c = 3.0 \times 10^{-6}$ | |
| Young's modulus $(GPa)$ | 308 | 201 | 68 |
| Acoustic velocity $(m/s)$ | 10,127 | 5,700 | 3,900 |
| Index of refraction | 2.08 | 2.01 | 2.4 |
| Band gap $(eV)$ | 6.2 | 3.4 | 2.67 |
| Resistivity $(\Omega cm)$ | $1 \times 10^{11}$ | $1 \times 10^7$ | $1 \times 10^9$ |

On the other hand, AlN has a low piezoelectric coefficient, but it has widespread applications because it has lower dielectric loss tangent, large bandgap, and large breakdown voltage. AlN also has a higher signal to noise ratio. AlN stands out from other materials as it has high-temperature stability and the strong polarity of their crystalline structure allows for polar growth and a stable piezoelectric response with time, whereas ferroelectrics always risk depoling. In addition, integration and process compatibility with the rest of the device is less difficult using the relatively simple wurtzite materials [1,6–8].

The comparative study in Figure 1.1 depicts the materials that satisfy the critical coupling criterion[2,9]. It neglects the influence of geometry. Materials with higher Q i.e. low damping provide more power and hence the coupling cannot be used as a dominant selection factor. Despite the lower $k_{31}$ factor of AlN compared to PZT, PMNPT or PZTPT materials, AlN provides sufficient coupling and could even produce more power because of its higher Q. Thus, choosing high-quality materials appears to be more of a concern than selecting those with high coupling.

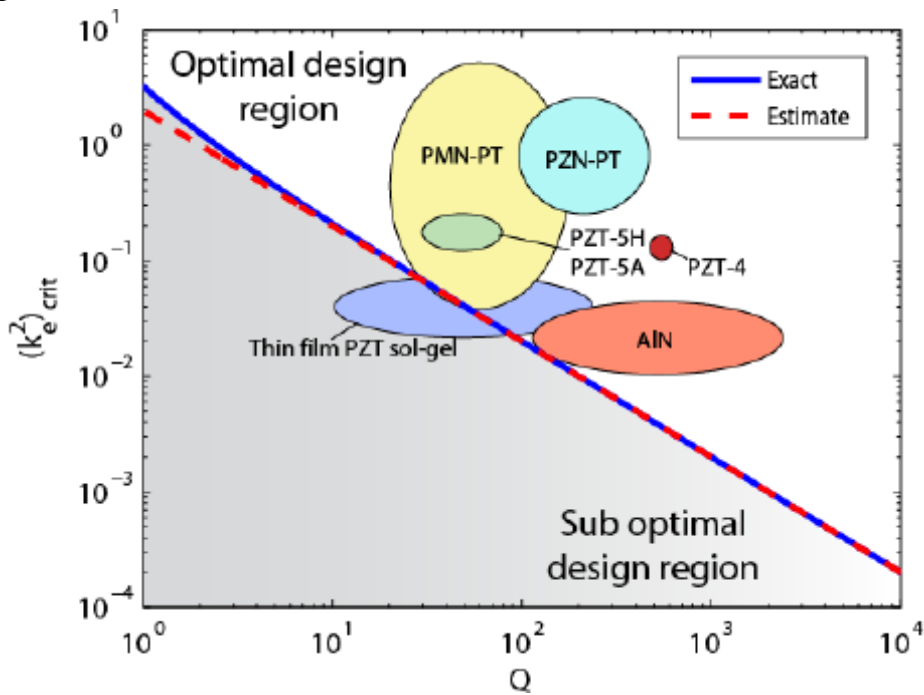

**Figure 1.1:** Piezoelectric materials comparison relatively to the minimal coupling criterion (Q=1/2ζm)

AlN is proven to be a very promising material for PiezoMEMS devices. In this work, we will focus upon the deposition of AlN on the Metal electrode by a sputtering technique. Also, we will address the challenges that need to be overcome to standardize AlN PiezoMEMS device fabrication. The challenge requires a start–to–finish effort, from the deposition and characterization of AlN films through the bulk micromachining of the devices to finally the device testing and analysis. To this day, there remain very few reported literatures that address all fabrication difficulties with proper characterization.

# CHAPTER 2
# Material Development

## 2.1 Introduction

### 2.1.1 Aluminum Nitride properties

AlN has a hexagonal wurtzite structure with a 6-mm point group and P6$_3$mc space group [10]. Its lattice constant a= 0.3110 nm and c= 0.4980 nm. In the crystalline lattice structure of AlN, Al atom is surrounded by four N atoms, forming a distorted tetrahedron with three Al-N(i) (i = 1,2,3) bonds named B$_1$ and one Al-N$_0$ bond in the direction of the c-axis, named B$_2$. The bond lengths of B$_1$ and B$_2$ are 0.1885 and 0.1917 nm, respectively. The bond angle N$_0$-Al-N$_1$ is 107.7° and that for N$_1$-Al-N$_2$ is 110.5°. Similarly, a tetrahedron is formed with the N atom as the center. Two tetrahedrons form a triangular prism with the C$_{3v}$ symmetry. In the AlN crystalline cell, the atoms of Al and N form four sp$^3$-hybridized orbits. The Al atom has three semi-full orbits and one empty orbit; the N atom has three semi-full orbits and one full orbit. The bond B$_2$ is formed by the coupling of the Al empty orbit and the N full orbit. Consequently, the ionic character of the B$_2$ bond is greater. The bond energy of B$_2$ is relatively smaller than that of the three equivalent B$_1$ bonds and is easy to break so that the energy required for sputtering particles to be deposited in the direction of the c-axis is great. From Fig. 2.1, it can be seen that the plane (100) is composed of bond B$_1$, while planes (002) and (101) consist of the bonds B$_1$ and B$_2$ together.

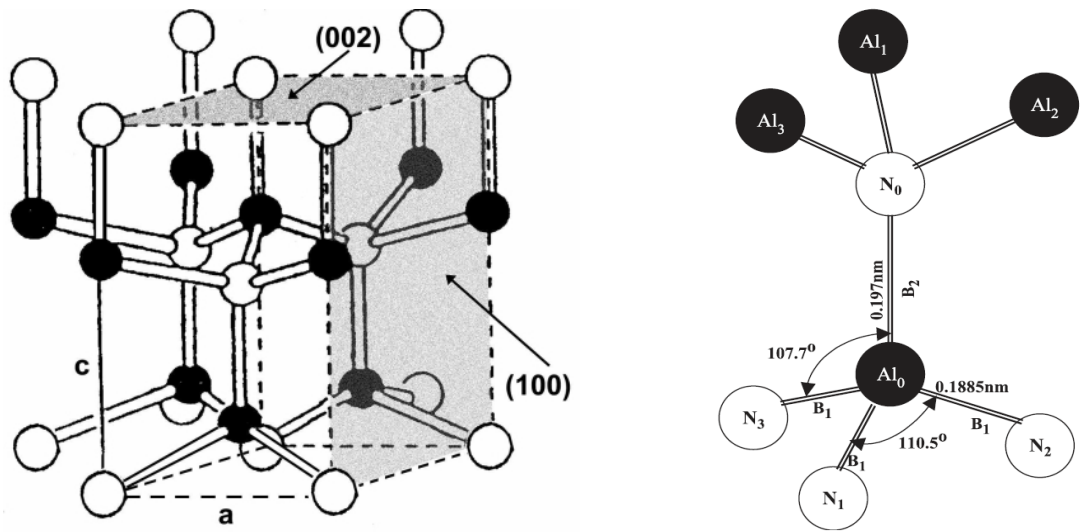

**Figure 2.1:** The crystal structure of AlN.

When preparing AlN films using the reactive magnetron sputtering method, the Al atoms sputtering from the surface of the Al target and reacting with activated N atoms or ions will form small AlN clusters. Combined with the time-of-flight mass spectra (TOF-MS), atom clusters with various sizes have been observed by Chu et al. [11].

The relationship between the mean free path ($\lambda$) of the sputtering particles and the distance (D) affects the orientation of the films. When $\lambda > D$, more sputtering particles reach the substrate directly without collision and the energy of the sputtering particles is greater, which is beneficial for the formation of bond $B_2$, and consequently, the growth rate of the (002) plane is faster. When $\lambda < D$, most of the sputtering particles will make collisions once or many times before reaching the substrate, thus decreasing the energy of the particles or forming small AlN clusters. In this case, the growth rate of the (100) plane is higher.

## 2.2 Requirements for AlN-PiezoMEMS

AlN is widely used in HEMT device stack as an interfacial layer between AlGaN and Silicon. In HEMT, AlN is used as a structural layer to reduce lattice mismatch between substrate and GaN. However, for PiezoMEMS devices, we need piezoelectric properties of AlN. Fig.2.2 shows effect of the crystalline quality of AlN i.e. Rocking curve FWHM of AlN (002) on piezoelectric properties like coupling coefficient ($k^2$) and piezoelectric constant ($d_{33}$). As we approach FWHM around 4°, AlN thin film achieves 99% of $k^2$ and $d_{33}$ value that of single-crystal bulk material[12,13].

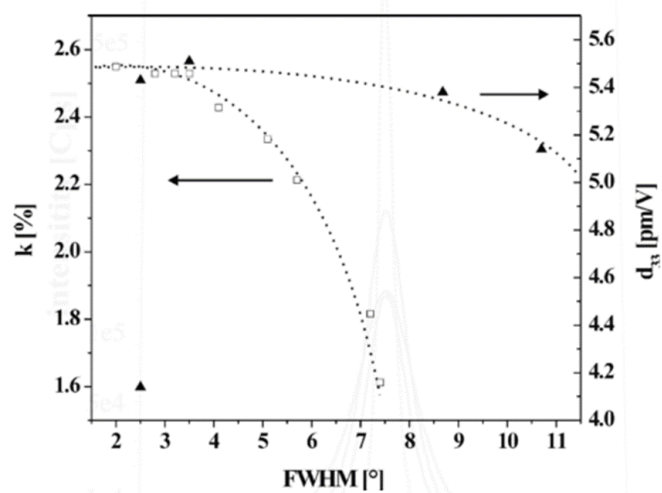

**Figure 2.2:** Comparison of the dependency of the coupling factor *k* in % and of the effective piezoelectric coefficient $d_{33\text{eff}}$ in pm/V on the FWHM in degree.

By keeping this in mind, we can define requirements for PiezoMEMS devices as follow:

1) AlN (002) rocking curve FWHM should be below 4°.

2) The roughness of film should be below 4nm for better performance.

3) Film thickness should be ~1μm with minimum residual stress.

4) The film should be deposited on the metal bottom electrode.

5) SOI substrate is favored.

6) Deposition technique should be cost-effective and scalable.

## 2.3 comparison of various deposition methods

Many different methods have been used to prepare AlN films. These include chemical vapour deposition (CVD), plasma-enhanced CVD (PECVD), filtered arc vacuum arc (FAVC), molecular beam deposition (MBE), hydride vapour phase epitaxy (HVPE), pulsed laser deposition (PLD), and sputtering. Of these technologies, MBE can grow a single-crystal epitaxial AlN film with other advantages which include precise control over the deposition parameters, atomic-scale control of film thickness and in situ diagnostic capabilities. However, it has limitations of low growth rate, expensive instrument setup and a high process temperature from 800 to 1000ºC. Unfortunately, this results in thermal damage of the AlN layers during deposition as well as the substrate, depending on the material. CVD technology including metal-organic CVD (MOCVD) and PECVD is also of great interest for AlN film growth because it not only gives rise to high-quality films but also is applicable to large-scale production. However, its high process temperature (about 500 to 1000°C) may be inappropriate for CMOS-compatible processes and this causes large thermal stresses in the films, which potentially restricts the choice of substrate. The main advantages of PLD are its ability to create high-energy source particles, permitting high-quality film growth at potentially low substrate temperatures (typically ranging from 200 to 800°C) in high ambient gas pressures in the $10^{-5}$–$10^{-1}$ Torr range. But the disadvantages of PLD are its limited deposition size and uniformity.

On the other hand, Reactive Magnetron sputtering offers low processing temperature, high deposition rate, and good texture and piezoelectric property. Thus, it is the most popular deposition technique for implementation of MEMS resonators for industrial and consumer electronic applications. Thus, we decided to deposit AlN by Magnetron sputtering.

## 2.4 Reactive Magnetron sputtering

### 2.4.1 Introduction

Reactive Sputtering is a process where a target of one chemical composition (here elemental Al) is sputtered in the presence of a gas or a mixture of gasses (Ar + $N_2$) that will react with the target material in the plasma and the thin film is formed of the reaction product (compound AlN). This technique can be performed with the DC as well as, the RF supply.

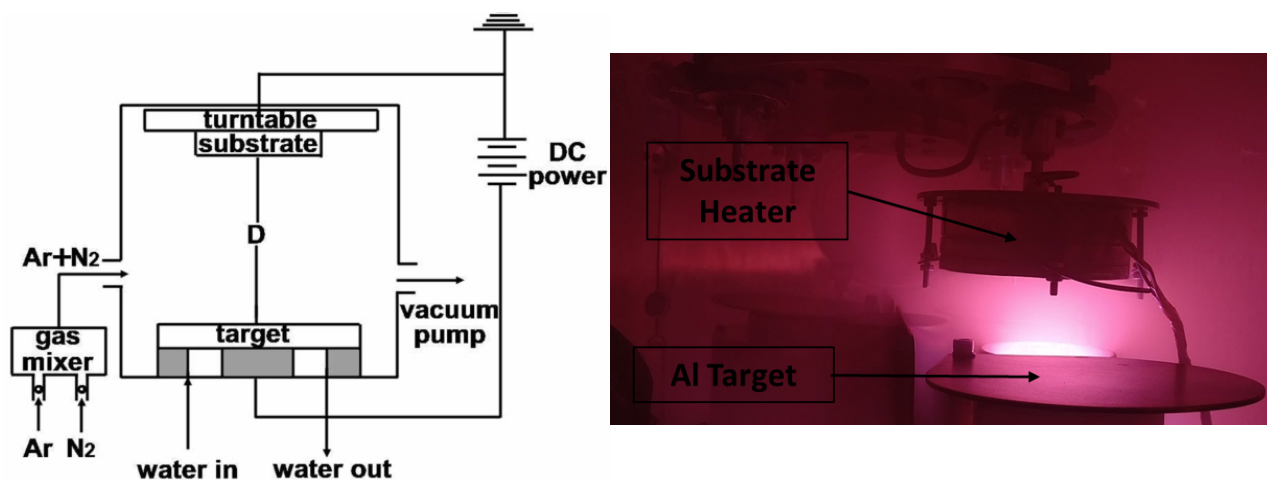

**Figure 2.3:** Schematics of AlN sputter tool deposition

#### 2.4.1.1 Parameter space

The parameters which are normally considered to controls the quality of AlN thin films deposited by reactive sputtering are as following:

Substrate

Layer thickness

Substrate temperature

Substrate bias

Base pressure

Target–substrate distance

Deposition rate

Gas flow ratio (Ar/$N_2$)

Discharge power

Process pressure

Out of these, Discharge power, Substrate, Substrate temperature, and Gas flow ratio (Ar/N$_2$) plays a significant role in the degree of c-axis orientation. Therefore, in this study, we will focus on the effect of these parameters on AlN thin film quality.

### 2.4.1.2 Power Mode

The sputtering tool has three types of power delivery system.

1) DC magnetron sputtering: In this, constant negative DC bias is applied to target. The major advantages of DC are that it is easy to control and cost-effective.

2) RF magnetron sputtering: RF Sputtering alternates the electrical charge at Radio Frequency so as to prevent a charge build-up on the target or coating material. In some cases, we can apply RF bias to the substrate to increase ion bombardment energy. Which enhances the crystal quality of the film.

3) Pulsed DC magnetron sputtering: During the pulsed-DC process, a negative bias is applied at frequencies ranging from a few to several hundred kHz; between pulses, a positive 'reverse' bias is applied to remove any built-up charge that has accumulated on the surface of the target. Advantage of Pulsed DC over RF is that it has a higher deposition rate and also the efficiency of deposition is high.

$$\%Duty\ Cycle = \frac{(Pulse\ ON)}{(Pulse\ ON) + (Pulse\ OFF)} \times 100$$

## 2.4.2 Effect of Power

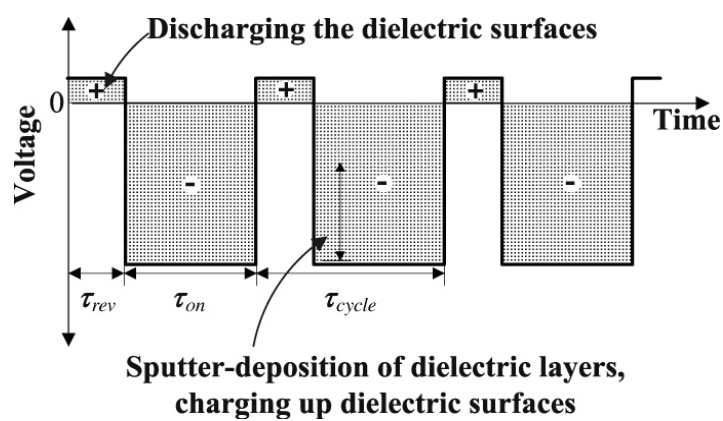

## 2.4.2.1 Power Mode

To study the effect of the power delivery system, the Experiment was done with keeping all other parameters constant and just varying mode of power delivery. Parameters used to study the effect of power mode are given in Table 2.1 B. residual stress analysis was done by KSA MOS UltraScan. In addition, deposition Rate was calculated on the basis of the deposition time and the thickness of the sample measured in SEM.

Table 2.1: Deposition parameters for various deposition conditions

| Parameters | (A) Pt | (B) Power Mode study | (C) Power Level study | (D) Pt substrate type study |
|---|---|---|---|---|
| Sputtering type | DC | DC, DC pulsed, RF | DC pulsed (150kHz 10%duty) | DC pulsed (150kHz 10%duty) |
| Sputtering Power | 17W | 250W | 250,400,500,600W | 250W |
| Target | Pt | Al (99.999%) | Al (99.999%) | Al (99.999%) |
| Target-Substrate Distance | 7.5cm | 7.5cm | 7.5cm | 7.5cm |
| Sputtering Temperature | RT | RT | RT | $500^0$C |
| Deposition Pressure | $2\times10^{-3}$Torr | $3\times10^{-3}$Torr | $3\times10^{-3}$Torr | $3\times10^{-3}$Torr |
| Atmosphere | Ar | $N_2$:Ar=3:2 | $N_2$:Ar=3:2 | $N_2$:Ar=3:2 |
| Substrate Rotation | No | No | 10rpm | No |
| Deposition Rate | 7.5nm/min | --- | --- | 8nm/min |

It was observed that DC and Pulsed DC had almost the same deposition rate. However, RF showed a decrease in the deposition rate. This may be attributed to the fact that in RF, during the positive cycle, the positive ions are accelerated to the surface of the target and sputter it.

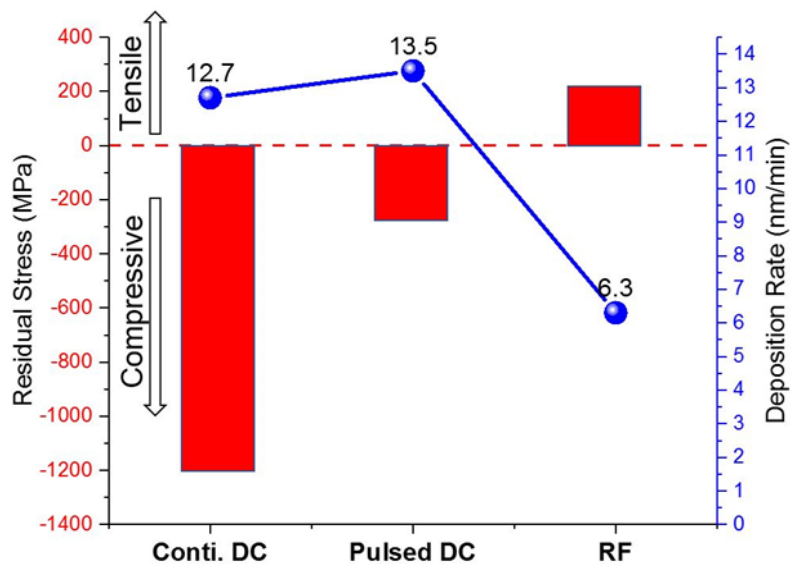

Figure 2.4: Effect of power mode on deposition rate and residual stress of 800nm AlN thin film

While in the negative cycle, the charging positive ions on the surface of the target gets removed. So effective deposition rate is slower than DC sputtering.

The main benefit of pulsed DC over the other two modes is that it has a higher deposition rate than RF with relatively low residual compressive stress. Thus, Pulsed DC was chosen to be the mode of deposition for further runs.

### 2.4.2.2 Power

For device fabrication point of view, deposited AlN film should have minimum residual stress. So pulsed DC mode was chosen to study the effect of power on the deposition rate and roughness. Parameters used to study the effect of power level are given in Table 2.1 C.

Roughness was measured using icon AFM. It was observed that, as we increase power, roughness value decreased from 4.06nm at 250W to 2.37nm at 500W and above that, it increased again. This effect can be explained with the help of adatom energy and adatom mobility. As we increase power, adatom energy increases and due to increased mobility and atoms finds a stable position on the substrate and roughness decreases. But as we increase power further, adatoms have sufficient energy but due to the high flux of atoms, it doesn't have enough time to go and sit in stable lattice position and leads to the rough surface.

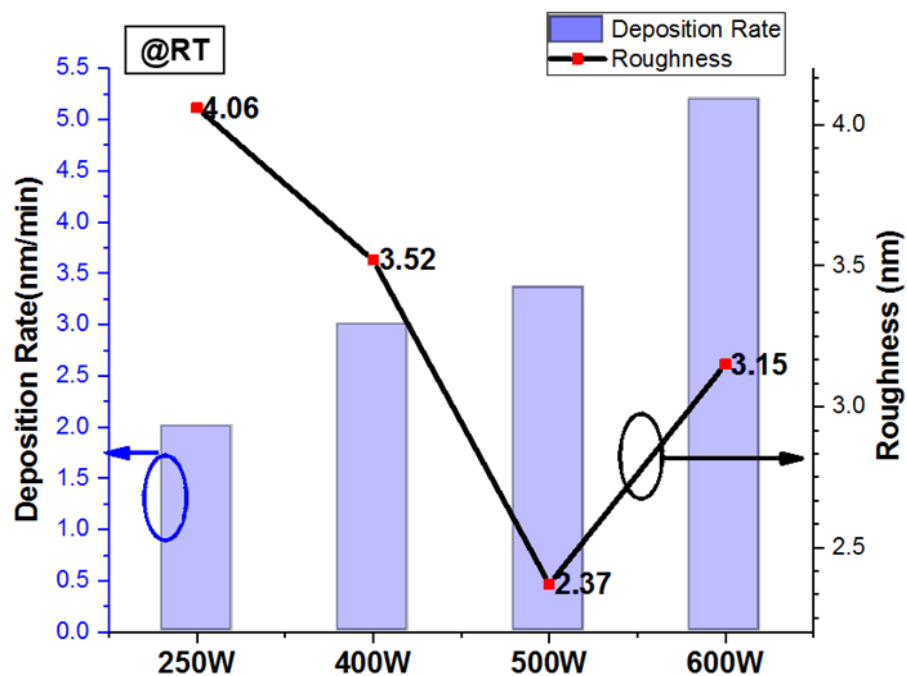

**Figure 2.5:** Effect of power level on deposition rate and roughness of AlN thin film.

from deposition rate data it is evident that as we increase deposition power, deposition rate increases proportionally. Since the increase in power causes more ion energy and which in turn causes higher bombardment of target and leads to a higher deposition rate.

### 2.4.2.3 Grain size comparison

Fig.2.6 shows the grain structure of AlN at different powers (phase image). We can clearly see that at lower power, grain size distribution is not uniform and smaller grain size (<25nm) dominates at lower powers i.e. 250W and 400W. But at higher power, grain size is fairly uniform with average grain size around 50nm. The dominance of small grains at lower power is because, at these power levels, the substrate doesn't have enough adatom mobility.

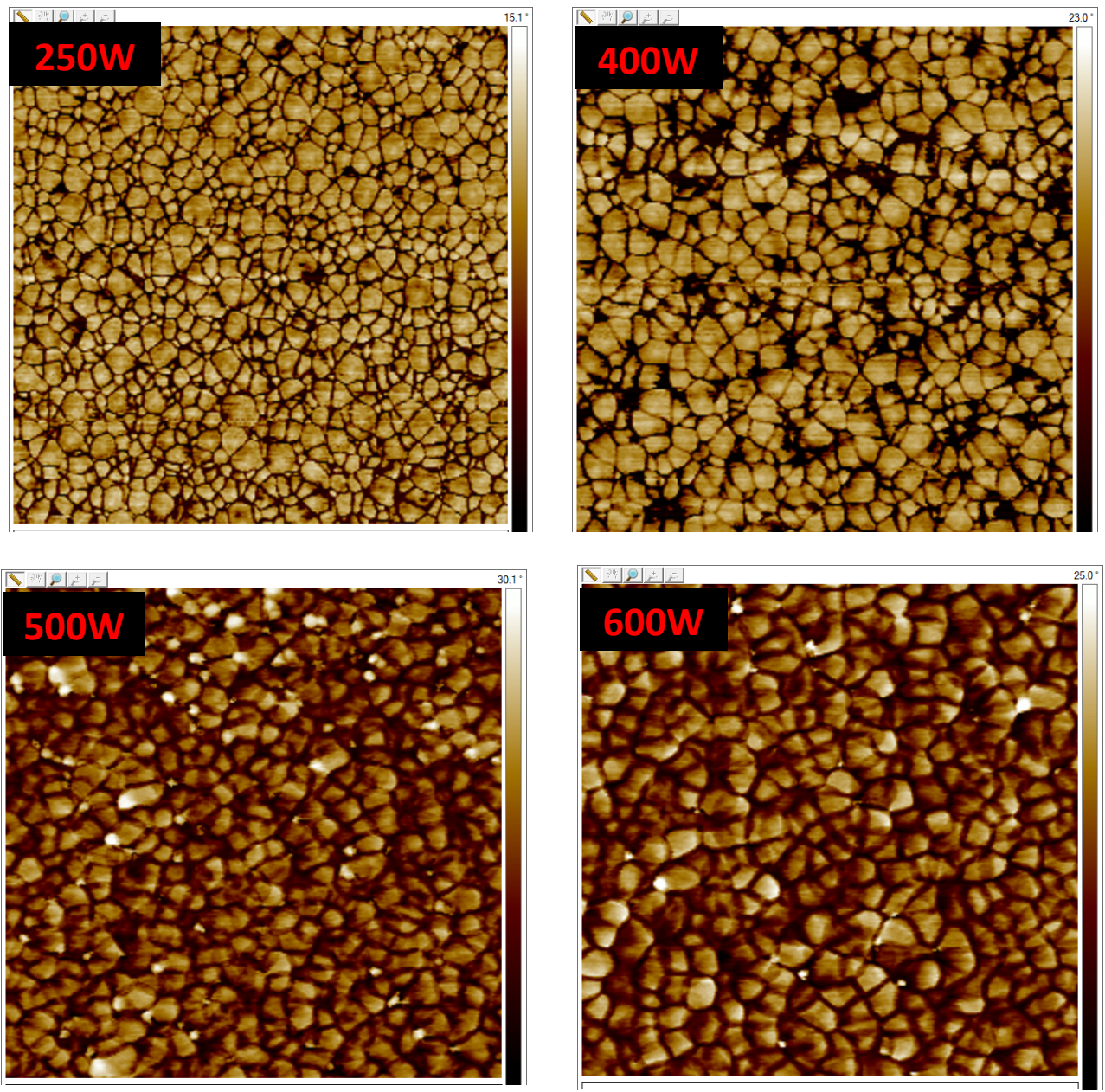

**Figure 2.6:** Comparison of deposited AlN grain size distribution with different power levels. Size: 1×1μm²).

### 2.4.3: Effect of the substrate stack

Another important parameter that governs AlN crystal quality is the bottom substrate. Therefore, a study was carried out on three different stacks to study the effect of bottom substrate on AlN film quality. Description of the stacks used is given below:

**Stack A:** In this stack, 150 nm Pt was deposited by sputtering on ALD deposited $TiO_2$ (Table 2.1 A). The reason to use ALD $TiO_2$ instead of Ti from sputtering is that sputtered Ti is porous and after annealing to $850^0C$, Pt diffuses through Ti and reacts with Si to form Pt-Si. According to metallographic studies on alloys annealed at $850^0C$, the solid solubility of Si in (Pt) is about 1.4 atom% [14]. Therefore, to avoid this, we used a conformal coating of $TiO_2$ deposited in ALD, which effectively improved the Pt crystal quality.

**Stack B:** Kamohara et al. [15] showed that the use of AlN interlayer below the metal electrode drastically improved AlN crystal quality. Therefore, we decided to use 150nm AlN layer grown on Si (111) wafer by MOCVD technique and deposit platinum with parameters explained in Table 2.1 A. But after doing characterization, it was observed that voids were formed on the film after annealing and $R_q$ (RMS roughness) came out to be ~650nm. The reason for such a bad quality AlN may be weak adhesion between sputtered Pt and atomically flat MOCVD AlN.

**Stack C:** To improve the orientation of Pt further, epitaxial silver deposited on Si (111) was used. Pt deposition parameters were kept the same as above (Table 2.1 A). The platinum showed epitaxial growth on Ag (111) with FWHM of $1.41^0$. SEM images also conformed epitaxial nature of Pt with hexagonal grain structure. After preparing these three stacks with different quality platinum, samples were mounted on the same holder to deposit ~800nm AlN by sputtering. The parameters used for this experiment are given in Table 2.1 D. Characterization of AlN was done using ICON AFM in tapping mode with $1\times1\mu m^2$ scan area. From roughness data, it was observed that the AlN roughness was minimum ($R_q$=1.93nm) on stack C and maximum on stack B ($R_q$=~650nm). The very high roughness of AlN on Stack B can be attributed to very rough platinum underneath the sputtered AlN. The grain size of AlN was uniform in both stack A and C at around 50nm. But stack B showed a grain size of ~650nm because of the above-mentioned reasons.

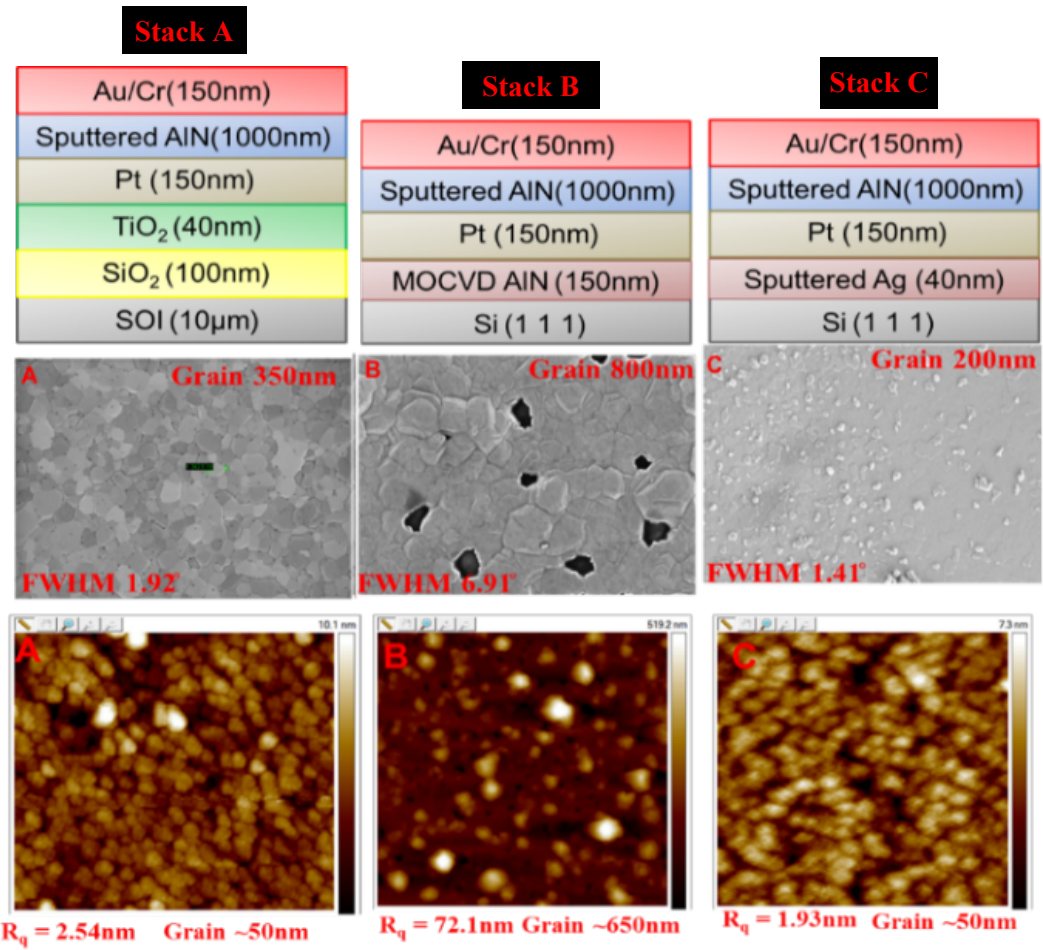

**Figure 2.7.** Comparison of quality of platinum (Row 2) and AlN (Row 3) deposited on that particular stack (images in Row 2 were captured in SEM with magnification 50,000×. images in Row 3 were captured using AFM. Image size: 1×1μm$^2$)

## 2.5 Characterization of the optimized thin-film stack

After completion of the optimization study, we achieved the best quality platinum and AlN. The rocking curve full-width at half-maximum (FWHM) of Pt (111) was measured to be 3° for thickness 150nm and that of AlN (002) was measured to be 4.5° for AlN of thickness ~800nm (Fig. 2.8-2.10). The rocking curve FWHM value depends on the thickness. Although these values are slightly greater than the best-reported value [16], K. Tonisch et al. [13] showed that for AlN FWHM below 5°, the piezoelectric coefficient $d_{33}$ achieves almost single crystal value.

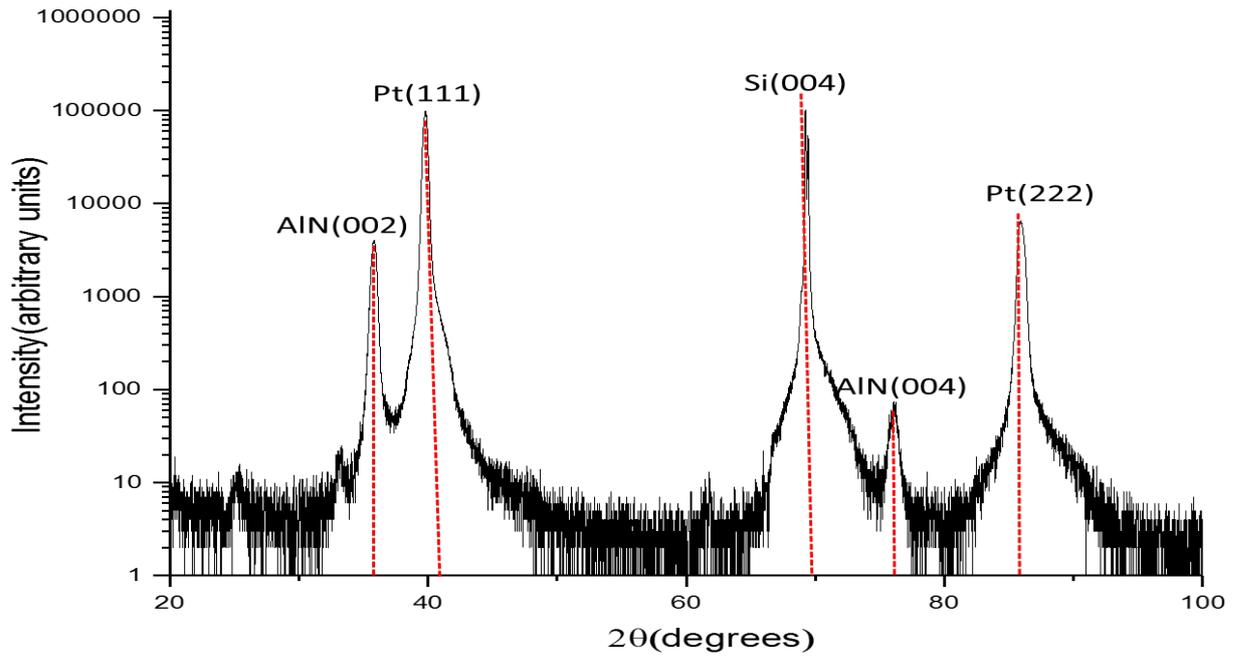

**Figure 2.8:** θ-2θ scan of sputter deposited thin film showing c-axis oriented (002) AlN thin films.

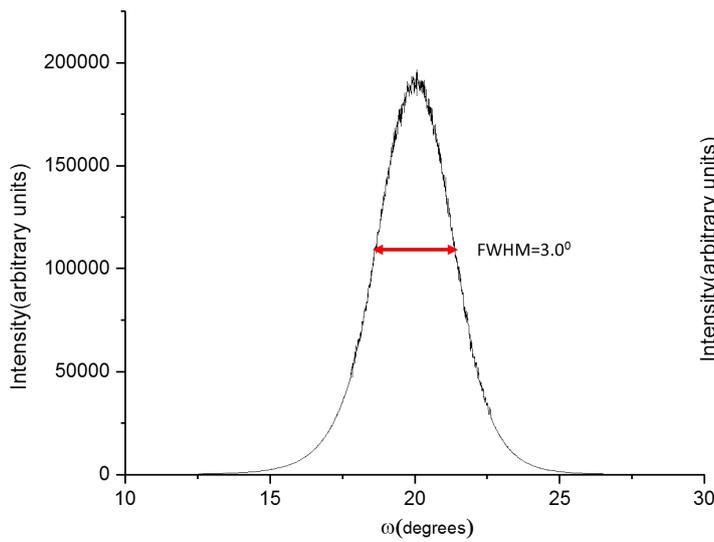

**Figure 2.9:** ω scan of sputter deposited Pt film showing rocking curve FWHM of Pt (111) =3.0°

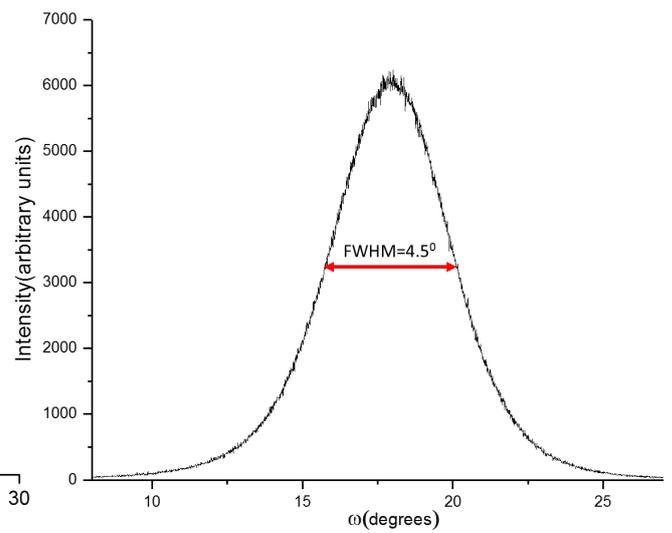

**Figure 2.10:** ω scan of sputter deposited AlN film showing rocking curve FWHM of AlN (002) =4.5°

It is evident from the SEM cross-section image that the sputter-deposited AlN film has a columnar grain structure. Similarly, the AFM micrograph indicates hexagonal grain shape of wurtzite AlN crystal (See figure 2.11, 2.12).

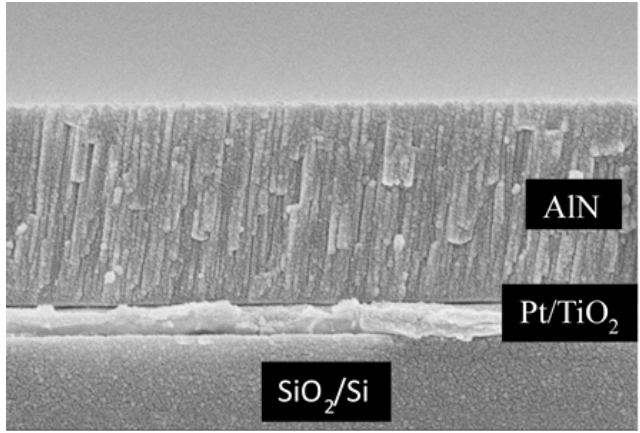

**Figure 2.11:** Cross-section SEM micrograph of sputter deposited Pt/AlN stack showing columnar grain growth

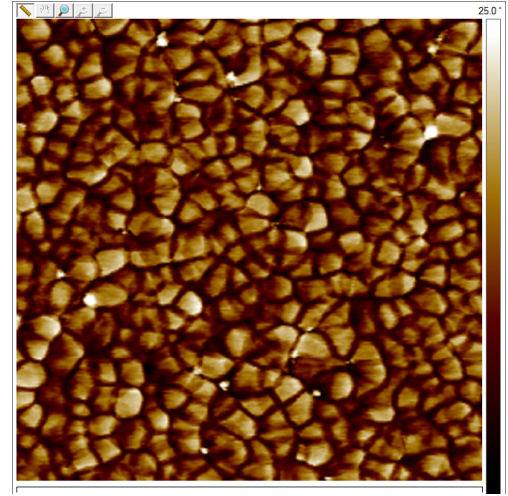

**Figure 2.12:** Top side AFM micrograph of sputter deposited Pt/AlN stack showing hexagonal grain growth.

## 2.6 Growth Mechanism of Aluminum Nitride Films

To study the growth mechanism of sputter-deposited AlN, we analyzed the data of AlN stack using cross-section TEM.

### 2.6.1 HR-TEM Analysis

It was observed that at MOCVD-Sputtered AlN interface, Sputtered AlN grains are oriented in a random manner. Which can be seen in the diffraction pattern also. Periodically arrange dots represent epitaxial MOCVD AlN and ring pattern indicate random nature of sputtered AlN. But as we go away from MOCVD-sputtered interface and toward the sputtered edge, grains get oriented along (002). This phenomenon was further studied in detail.

From figure 2.14, it is evident that during the initial stage of deposition, initial nucleation occurs with random orientation. But as crystal growth continues, crystal orientation with fastest growth rates outgrow unpreferred, slower orientations and the film coalesces into a polycrystalline film with a preferred orientation.

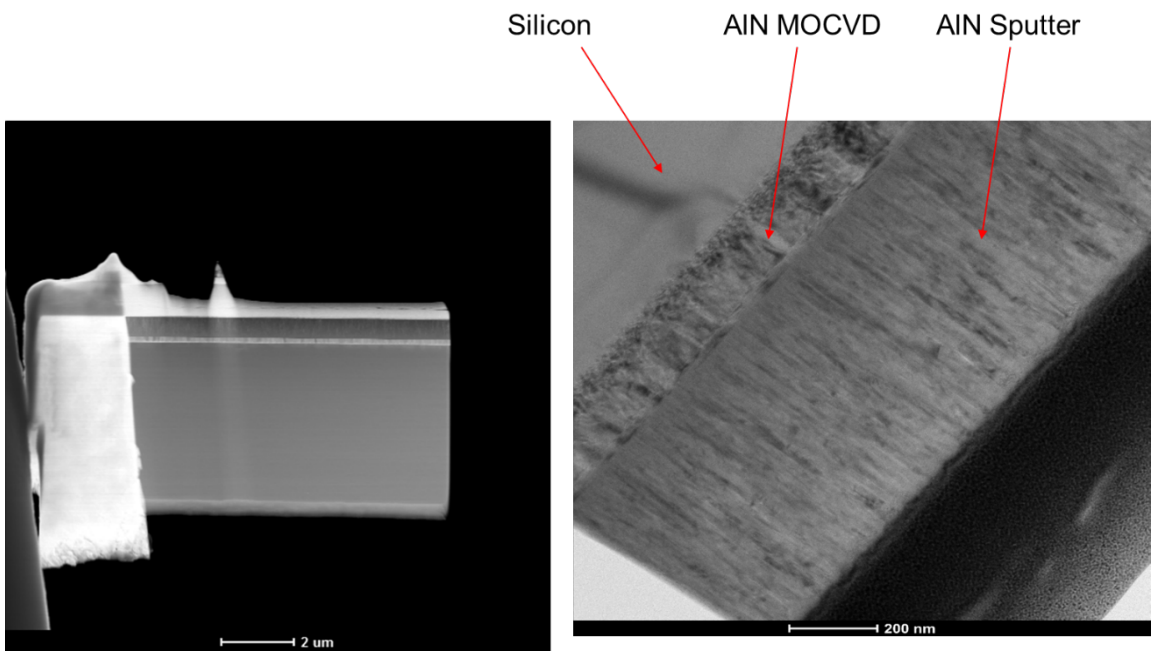

**Figure 2.13:** AlN cross-section TEM at low Magnification

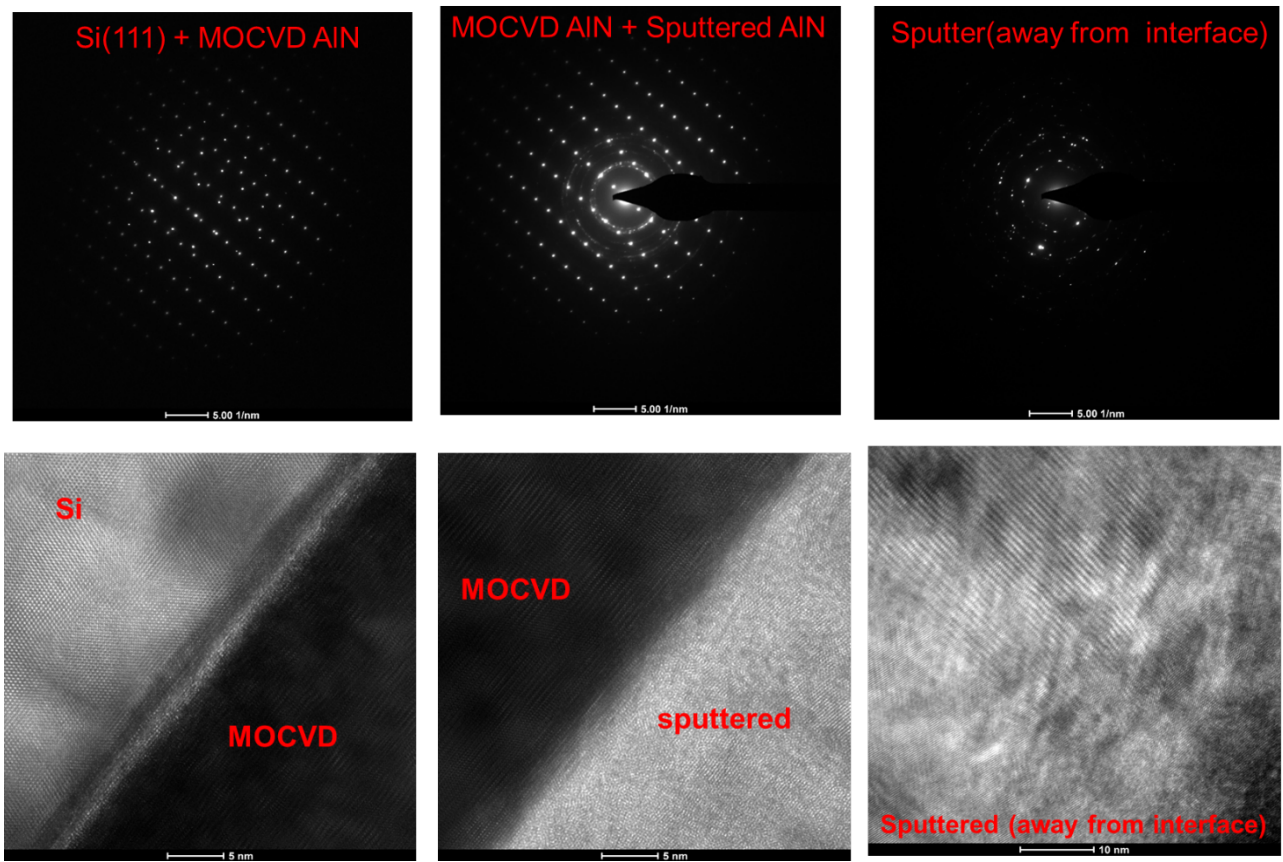

**Figure 2.14:** AlN cross-section HR-TEM showing Si-MOCVD AlN and MOCVD AlN/Sputtered AlN interface

The Low–temperature reactive sputtering of AlN occurs at Ts/ Tm values between 0.1 and 0.2 and pressures well below 1 mbar. These conditions correspond to zone T growth [17](figure 2.15). This growth comprises of four stages which are illustrated in figure 2.16[5]. One thing is clear that these islands are not epitaxial with the substrate since the growth temperature is too low. Also, the substrate and the AlN film are not lattice-matched.

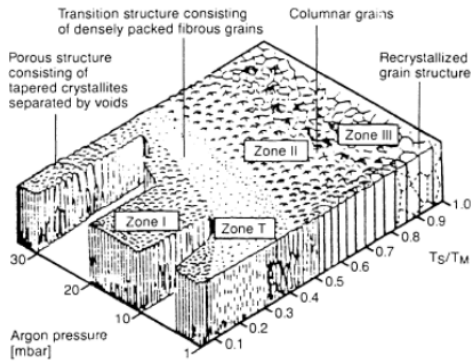

**Figure 2.15:** Thornton's structure zone model for sputtered thin films. The process conditions for low temperature AlN sputtering correlate to the microstructure of zone T.

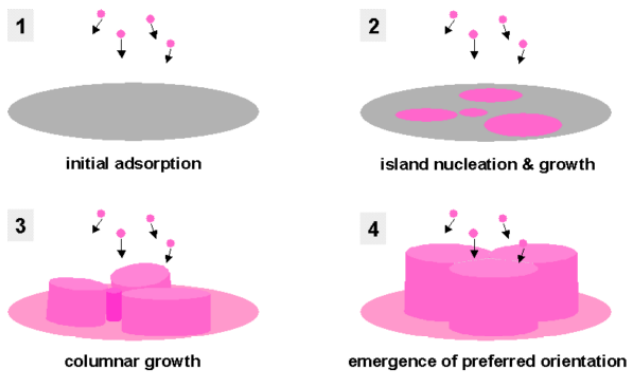

**Figure 2.16:** Growth mechanism of an AlN thin film. A four-stage process occurs during sputter deposition: 1) atoms and molecules sputtered from the target travel to the surface of the substrate and form nucleation sites. 2) Islands grow from the nucleation sites. 3) The islands coalesce into a film containing many crystallites with various orientations. 4) Crystal orientations with the fastest growth rates outgrow unpreferred, slower orientations and the film coalesces into a polycrystalline film with a preferred orientation.

Figure 2.17 shows cross-section HR-TEM micrograph of sputtered AlN with optimized parameters. From the diffraction pattern, it is evident that AlN has a columnar grain structure with preferable orientation in (002) direction.

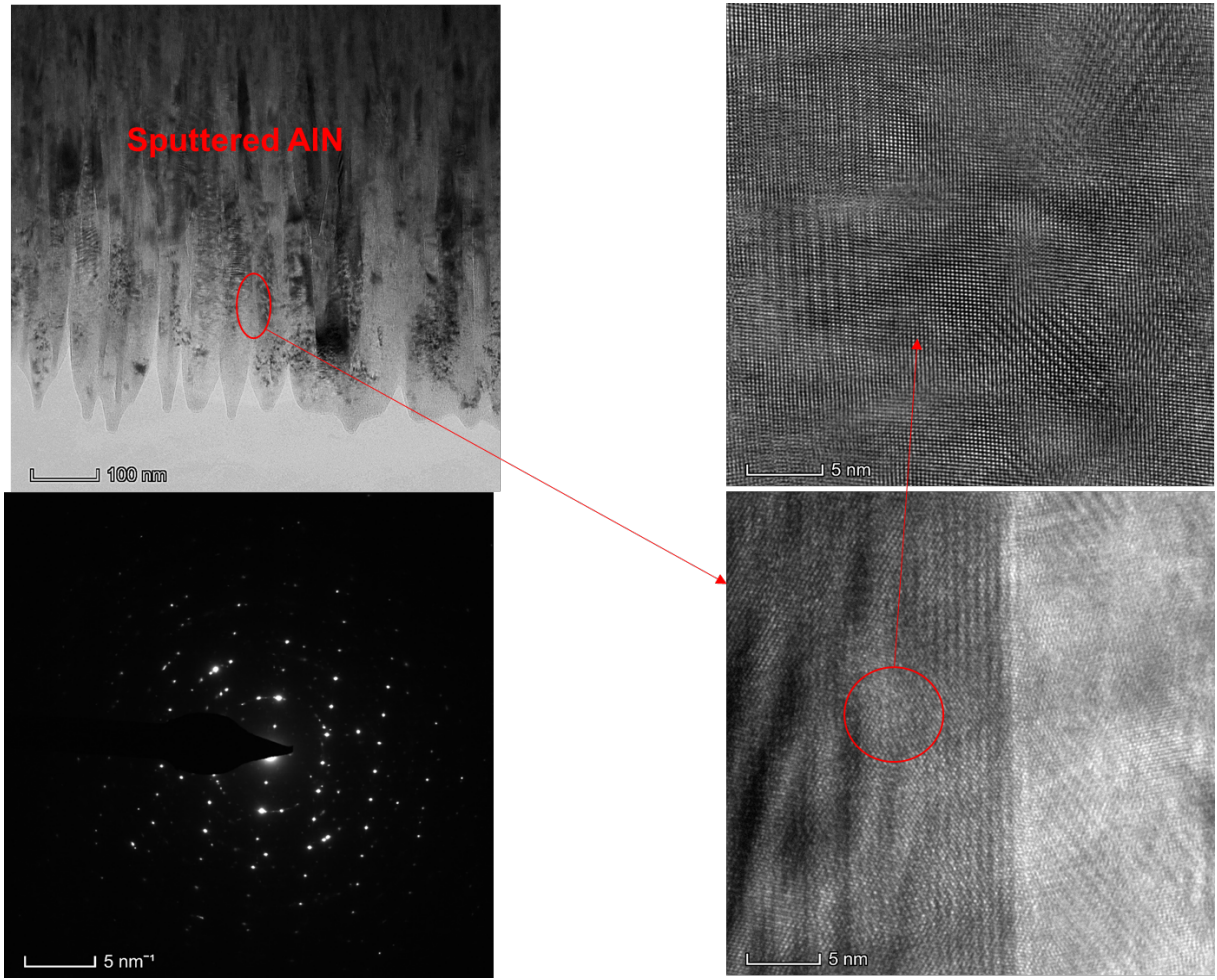

**Figure 2.17:** Sputtered AlN cross-section HR-TEM showing columnar grains of AlN oriented along c-axis (002).

### 2.6.2 TEM Elemental Analysis

The sample was used to characterize elemental distribution across the deposited layers (fig. 2.18). From EDS analysis, it was observed that there is increased in the concentration of oxygen at the interface of MOCVD-Sputtered AlN. Which means there were few monolayers of amorphous native oxide present on MOCVD surface. Because of which, initial nucleation sites of Sputtered AlN did not grow epitaxially.

This problem can be solved by precleaning of MOCVD AlN. So to do so, we used 1:4 $HCl:H_2O$ for 30 sec to remove any organic contaminations and then 1:50 $HF:H_2O$ for 30 sec to remove native oxide. Then cross-section TEM was done. But in EDS analysis we could still able to see oxygen traces at the MOCVD-Sputtered AlN interface. The reason for this may be after wet etching/cleaning the sample, it is not possible to load the sample immediately and start deposition. One way to solve this problem can be to apply a reverse bias to the substrate and do in-situ cleaning in the presence of Ar gas. This in-situ cleaning need to be studied further in detail(see fig. 2.19).

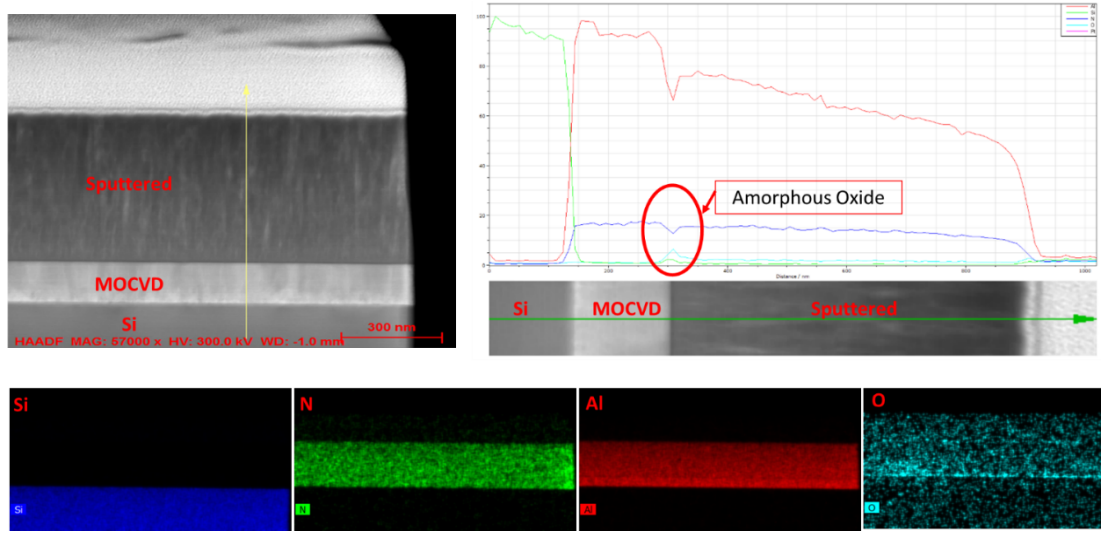

**Figure 2.18:** TEM elemental analysis of AlN film without surface wet etching

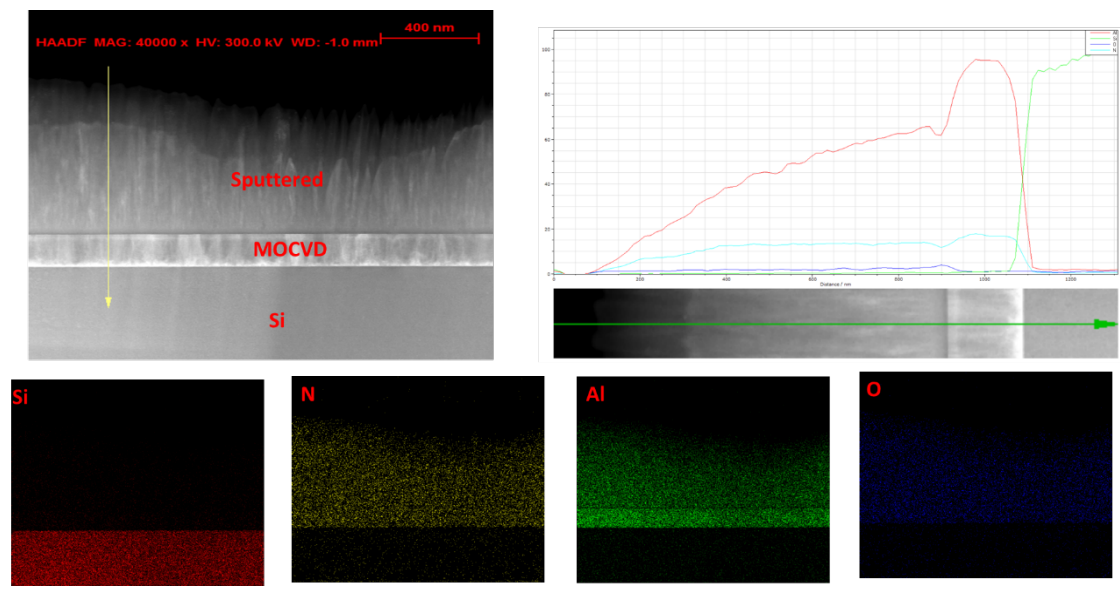

**Figure 2.19:** TEM elemental analysis of AlN film with surface wet etching

# CHAPTER 3
# AlN-Based PMUTs Fabrication

## 3.1 Introduction

The previous section describes how to control crystal properties of AlN to obtain good quality polycrystalline AlN thin films. Next step in the process is device fabrication. Here we show fabrication of Piezoelectric Micromachined Ultrasonic Transducers (PMUTs) with AlN as an active piezoelectric layer.

## 3.2 Process Flow

The Micro-fabrication of PMUTs is illustrated in the flow chart [18]:

**Table 3.1:** Stepwise process flow for PMUTs fabrication

| | Step Name/Description: | Unit Device cross section | Mask/Top view (if applicable) |
|---|---|---|---|
| 1 | Substrate cleaning/RCA | 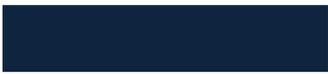 | |
| 2 | SiO$_2$ Layer Growth/First Nano Dry oxidation | 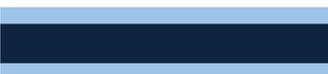 | |
| 3 | TiO$_2$ Layer Deposition/Atomic Layer Deposition | 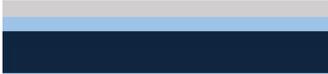 | |
| 3 | Platinum Layer Sputtering/Sputter coater 1 | 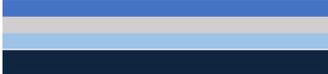 | |
| 4 | Platinum Annealing/ Tempress_Metal Anneal | 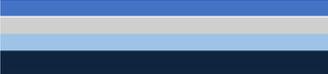 | |
| 5 | AlN Sputtering/ Sputter coater 2 | 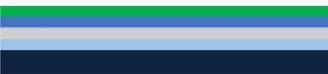 | |
| 6 | Top Electrode Optical lithography/EVG620 | 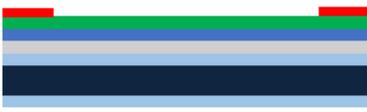 | 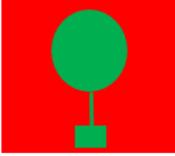 |
| 7 | Top Electrode Au/Cr Sputtering and Lift Off /Sputter coater 1 | 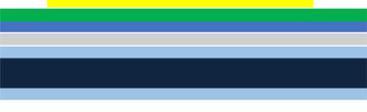 | 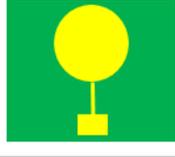 |
| 8 | AlN Etching/RIE-Cl | 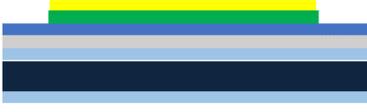 | 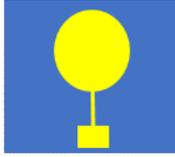 |
| 9 | Back Side Lithography/EVG620 | 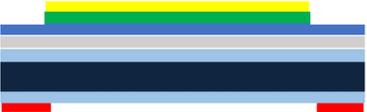 | 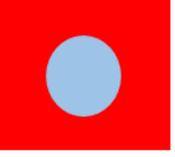 |
| 10 | Backside SiO$_2$ RIE and PR strip | 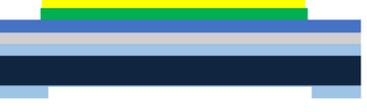 | 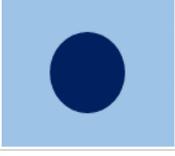 |
| 11 | Backside DRIE | 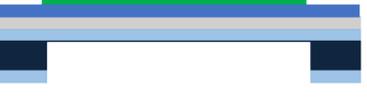 | |
| 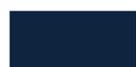 | SOI | 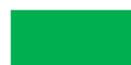 | AlN |
| 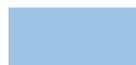 | SiO$_2$ | 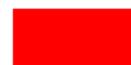 | Photoresist |
| 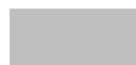 | TiO$_2$ | 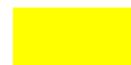 | Au/Cr |
| 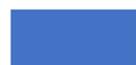 | Pt | | |

Following is a stepwise description of the fabrication process:

1. Substrate cleaning:

First of all, fresh SOI wafer needs to be cleaned before depositing thermal oxide in Level 1 wet bench (Bulk Process). RCA-1 cleaning and RCA-2 cleanings are carried out for the wafer for 10 mins each followed by a 30 seconds dip in buffer-HF solution to remove any organic contamination or native oxide from the wafer (Table 3.1-1).

2. Thermal Oxide deposition:

100nm $SiO_2$ layer is grown on both sides of the wafer by heating the clean wafer to 1100 °C for approximately 1.5 hrs (Table 3.1-2). The $SiO_2$ layer carries ~ 300 MPa compressive residual stress.

3. Bottom electrode sputtering;

a) $TiO_2$ Layer Deposition: 20nm $TiO_2$ layer is deposited on top of the oxide layer by using Atomic Layer Deposition (ALD) tool. The recipe used for this is 'TTIP+$H_2O$ $TiO_2$ deposition'. The $TiO_2$ deposited is polycrystalline Anatase in nature (Table 3.1-3).

b) Platinum layer deposition: 150nm Pt is deposited on the top of the $TiO_2$ layer by using DC magnetron sputtering at room temperature. To improve Pt crystal quality further, the wafer undergoes annealing process at 850°C for 3hr in $N_2$ atmosphere (Table 3.1-4). Sputtering parameter for Pt are mentioned in Table 3.2 A.

4. AlN Sputtering:

800nm thick AlN layer is deposited on the sample by following the process described in Table 3.2 B.

Table 3.2: Sputter deposition parameters for Pt and AlN

| Parameters | (A) Pt | (B) AlN |
|---|---|---|
| Sputtering type | DC | DC pulsed (150kHz 10%duty) |
| Sputtering Power | 17W | 250W |
| Target | Pt | Al (99.999%) |
| Target-Substrate Distance | 7.5cm | 7.5cm |
| Sputtering Temperature | RT | 500°C |
| Deposition Pressure | $2\times10^{-3}$Torr | $3\times10^{-3}$Torr |
| Atmosphere | Ar | $N_2$:Ar=3:2 |
| Substrate Rotation | No | No |
| Deposition Rate | 7.5nm/min | 8nm/min |

5. Top Electrode Lithography:

The AlN deposited wafer is spin-coated with photoresist (PR) AZ541E at 4000 RPM for 40 seconds. The PR coated wafer is soft-baked at 95°C for 1 min and then exposed to UV radiation with ~45 mJ/cm$^2$ energy through chrome mask with top electrode patterns using EVG-620 optical lithography tool. The wafer is then developed dipping it in AZ351B PR-Developer for 25 seconds (Table 3.1-6).

6. Top electrode sputtering:

30 nm Cr adhesion layer followed by 150 nm Au layer is deposited on the sample with a patterned photoresist layer. The sample with deposited Cr/Au film is soaked in acetone for 2 hrs followed by 5 mins of sonication in order to dissolve the PR and lift-off the Cr/Au layer from the sample leaving behind the patterned top electrode layer on the sample (Table 3.1-7).

7. AlN Etching:

To expose the bottom electrode, we need to etch the AlN layer in the selected area (Table 3.1-8). This is done by either a wet etching recipe or Dry etching recipe. Details of AlN etching are described in the subsequent section.

8. Back Side Lithography:

The top side processing is now complete. To do backside etching, we need to deposit 2μm SiO$_2$ hard mask by PECVD technique. Then the sample is coated with AZ4562 photoresists at 4000 RPM for 30 seconds to obtain ~6.5μm thick photoresist layer on the backside. Soft backing is done at 95°C for 1 min. The sample is loaded on to the EVG-620 system, aligned with the bottom side hard mask with the help of a bottom microscope and exposed to 110 mJ/cm$^2$ of UV radiation. Following this, the PR is developed using AZ351B developer for 70 seconds (Table 3.1-9).

8. Backside Oxide Etch:

The 2μm thick SiO$_2$ layer is etched using the reactive ion etching (RIE) method. This process uses CHF$_3$ gas (Flow Rate: 40sccm) and bombards the substrate with generated radicals at low pressure using plasma at 1500-Watt RF power and then PR ashing is done to remove AZ4562 (Table 3.1-10)

9. Deep Reactive Ion Etching (DRIE):

To release the diaphragm, we need to etch the backside silicon handle layer (525μm). For this, we use the DRIE technique to make high aspect ratio holes. In this step, the sample is loaded into a Deep Reactive Ion Etching system and exposed alternately to $SF_6$ and $C_4F_8$ gases under low pressure environment with high power plasma for approximately 20 mins to etch ~ 525 μm silicon through the patterns made on the backside of the wafer, landing on the buried oxide of the SOI wafer. The buried oxide is etched out using the same RIE process as described in step 10 for approximately 5 mins to obtain the final released PMUT structure (Table 3.1-11)

## 3.3 AlN Etching

To implement active elements made of AlN in the process flow of a MEMS device, an adequate patterning procedure is necessary. A straightforward approach is the use of wet etchants, such as phosphoric acid ($H_3PO_4$) or KOH, as these chemical products are well established in silicon micromachining. In addition, a wet etching process requires inexpensive equipment compared to a dry-etching technique.

After the successful deposition of AlN film, we also need to etch it from selective area to expose the bottom electrode for electrical contacts. Many literatures show that AlN can be etched by either a wet etching technique or Dry etching technique [19–22]. Both techniques have its advantages and disadvantages. Wet etching gives good selectivity between AlN and Platinum electrode underneath also etch rate is very high. But it is difficult to control and also suffers from a lack of repeatability. On the other hand, RIE is quite controllable and repeatable after optimization. However, the main hurdle in RIE is that after etching AlN, Pt bottom electrode gets exposed to reactive plasma and forms a non-volatile product which is difficult to remove from chamber wall and which hampers etch rate of AlN and which in turn causes lack of repeatability.

### 3.3.1 AlN wet Etching

To implement active elements made of AlN in the process flow of a MEMS device, an adequate patterning procedure is necessary. A straight-forward approach is the use of wet etchants, such as phosphoric acid ($H_3PO_4$) or KOH, as these chemical products are well-established in silicon micromachining. In addition, a wet etching process requires inexpensive equipment compared to a dry-etching technique. However, in this work focus is on the effect of KOH etchant temperature on AlN etch rate.

Gold top electrode was used as a mask for the wet chemical etching of the AlN layers in 10% KOH. The temperature of the etchant was varied between 90°C and 50°C. Etch rate was measured by SEM cross-section imaging.

### A) 10% KOH at 90°C

At 90°C, it was observed that, within 30sec, 1μm AlN got etched along with the thickness and around 3.2 μm was etched along the lateral direction. After 1min, lateral etching was increased to 4.2 μm. From this experiment, the etch rate along thickness was calculated to be ~2 μm/min and along the lateral direction to be ~4.2 μm/min. Therefore, this recipe was etching AlN in lateral direction 2 times faster than thickness direction. This lateral etching causes undercut and which may result in shorting of the top electrode with the bottom electrode.

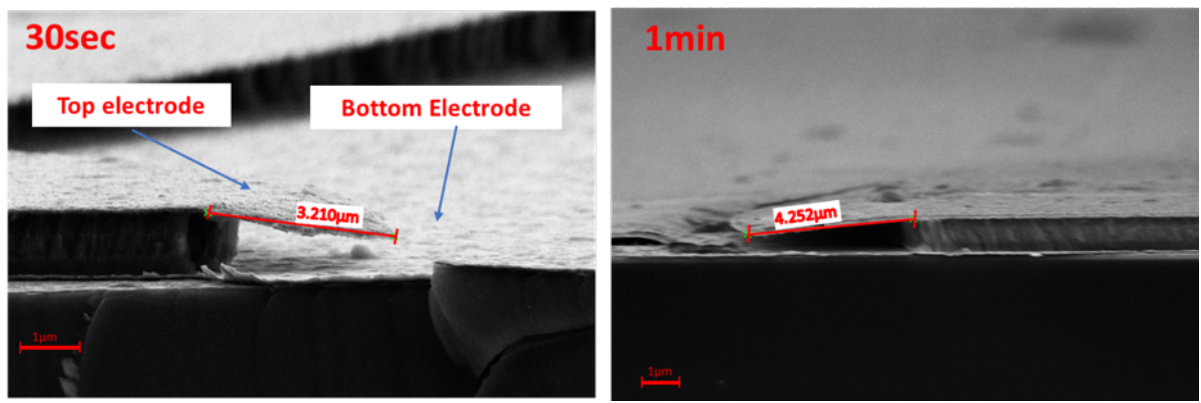

**Figure 3.1:** SEM micrographs of AlN etched by 10% KOH at 90°C showing undercutting

### B) 10% KOH at 50°C

To avoid the problem of undercutting and faster etch rate, we tried AlN etching using 10% KOH at 50°C. The etch rate follows the Arrhenius equation. As the temperature was reduced from 90°C to 50°C, etch rate reduced to ~300nm/min along thickness and ~70nm/min along the lateral direction. From this study, it can be concluded that temperature plays an important role in controlling etch anisotropy. For making good quality devices, we ideally need a vertical etch wall. i.e. no lateral etching at all. But this can not be achieved in polycrystalline AlN with wet etching recipe. But for all practical purpose, AlN etching at 50°C will work well.

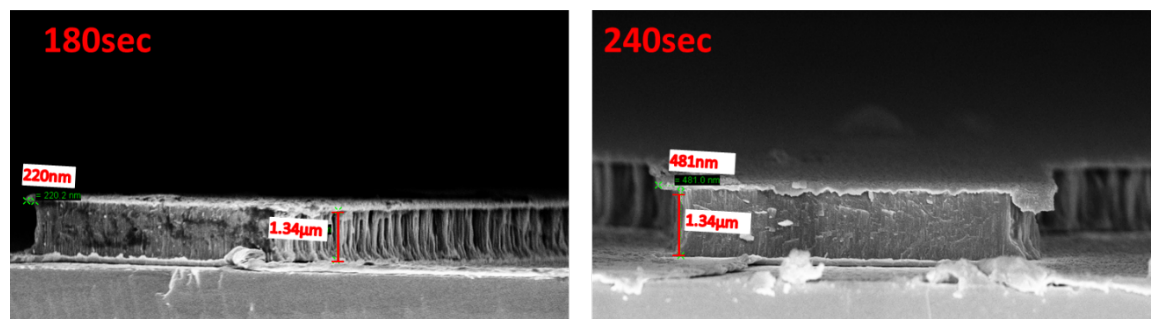

**Figure 3.2:** SEM micrographs of AlN etched by 10% KOH at 50°C showing no undercutting

As discussed earlier, it is difficult to control temperature and KOH concentration during etching. Due to this, we cannot gate repeatable results.

### 3.3.2 AlN Dry Etching by RIE

After facing the problem of repeatability in the wet etching process, we studied the AlN etching process by Inductively coupled plasma (ICP) RIE process. The etching sample in this experiment is polycrystalline AlN film with thickness ~800nm with root-mean-square (RMS) roughness 2.54nm measured by atomic force microscope (AFM). The grain size value is approximately 50 nm. Two types of PR were used. One was positive PR(AZ4562) with thickness ~1.5μm and second is negative PR(AZnOLF2020) with thickness ~6 μm. Chlorine gas was used in the etching process of AlN. The etching products are a series of Al‑Cl volatile compounds, such as $AlCl_3$, $Al_2Cl_6$, and other Al-Cl compounds.

#### A) RIE Recipe

The experiment was performed at 500W ICP power with different etch time. The etched step height was measured with the help of Bruker DekTek XT Surface Profilometer and Bruker Dimension Icon AFM to calculate selectivity between PR: AlN and AlN: Pt bottom electrode. We also varied ICP Power to check the etch rate.

#### B) RIE result and Discussion

Following observations were made during the experiment:

- Low power recipe gave moderate etch rate and good selectivity between AlN and PR.

- Negative PR (AZnOLF2020) was able to sustain at all powers without changing color. But Positive PR was becoming Brownish. And at higher power and time (1500W and 6min) surface became bubbly.

- It was observed that Ti/Pt (RT, W/O Annealed) was etching faster than $TiO_2$/Pt (850°C Annealed).

- SEM imaging showed AlN spikes.

- The selectivity of AlN: Pt was ~9:1 and AlN: PR(AZnOLF2020) was 0.5:1

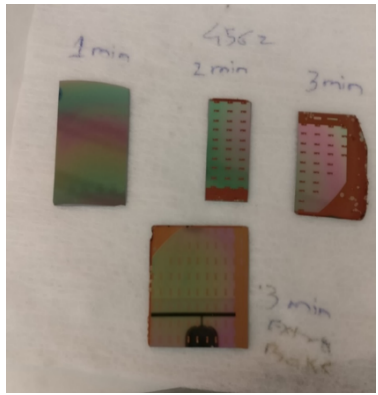

**Figure 3.3:** Samples after AlN etching with positive PR (AZ4562).PR becoming brownish in color. a) 1min b) 2min c) 3min d) 3min (extra 5min PR bake)

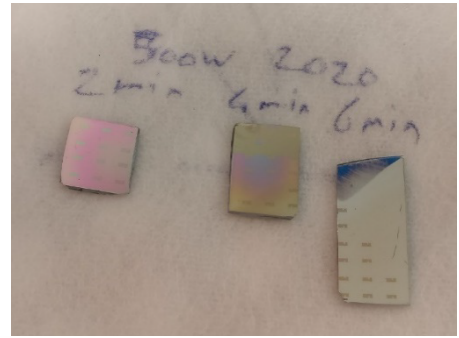

**Figure 3.4:** Samples after AlN etching (AZnOLF2020) a) 2min b) 4min c) 3min d) 6min

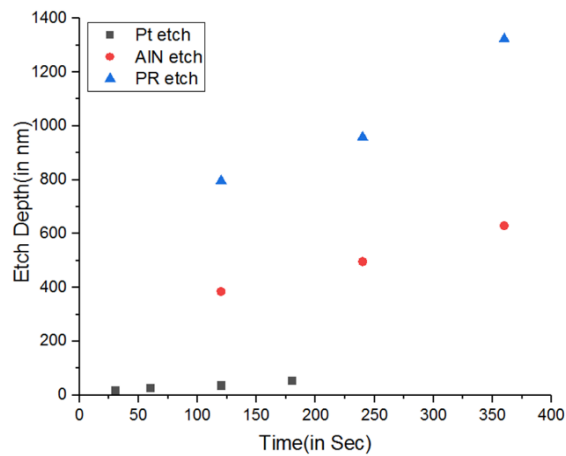

| Time | AlN:Pt | AlN:PR |
|------|--------|--------|
| 120  | 9.5:1  | 0.50:1 |
| 180  | 8.1:1  | 0.49:1 |

**Figure 3.5:** Selectivity of AlN with respect to Pt and PR (AZnOLF2020) in RIE

## C) RIE Issues

AlN etching by RIE is reported in many literatures [19–22]. But in any research facility usually takes up several runs to optimize it for a particular need. During optimization, we faced several problems. This study gave us better understanding of the problem, thus, leading to a better solution.

### a) Platinum Etching:

The important hurdle in the etching process was exposing of Pt bottom electrode to reactive plasma. As mention previously, Pt forms non-volatile compound after being etched out and this was causing serious problem to RIE-Cl tool (Etch rate deviation in other recipes). So one solution to this problem may be the use of Molybdenum (Mo) as the bottom electrode. Since Mo forms volatile products in RIE.

**b) AlN spikes:**

After etching AlN in RIE, we saw AlN spikes in the etched area. The reason for this etch anisotropy is due to the crystal orientation of AlN. Since different crystallographic planes etch at a different rate. To remove these spikes, we can do wet etching using 10% KOH at 50$^\circ$C for few seconds. This will ensure the removal of AlN spikes and also Pt will remain unetched.

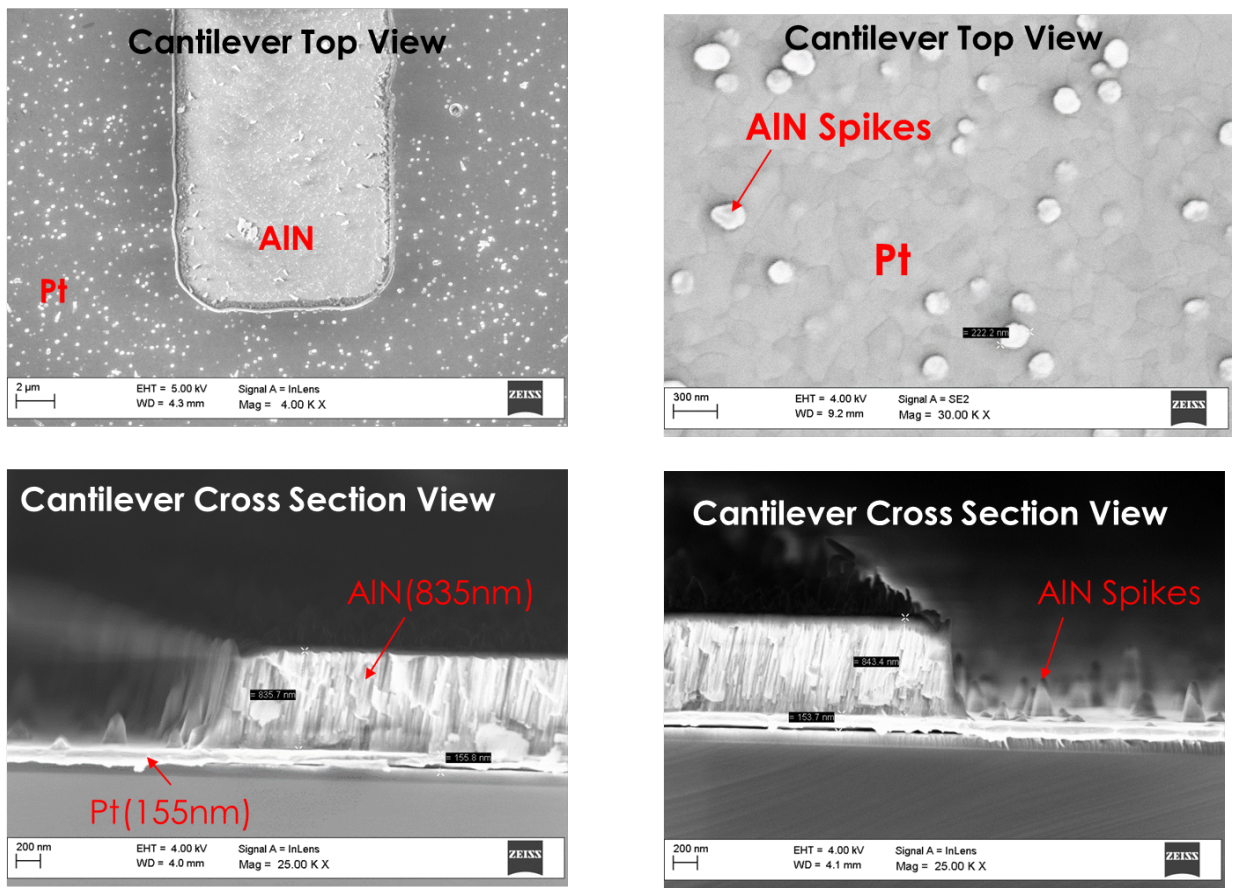

**Figure 3.7:** SEM micrograph of AlN RIE showing AlN spikes

## 3.4 Back Side DRIE

DRIE is the final and most critical step in device fabrication. This process is very sensitive to small variations in etching parameters and these variations in the process play a paramount role in the realization of the final structure.

### 3.4.1 Undercutting

The problem of undercutting was observed during fabrication of high fill factor PMUTs. The degree of undercut depends upon fill factor (ratio of etched portion to unetched portion) and time of etching. After etching silicon up to Buried Oxide (BOx) layer, Silicon started to etch in the lateral direction and this caused the merging of the released diaphragm.

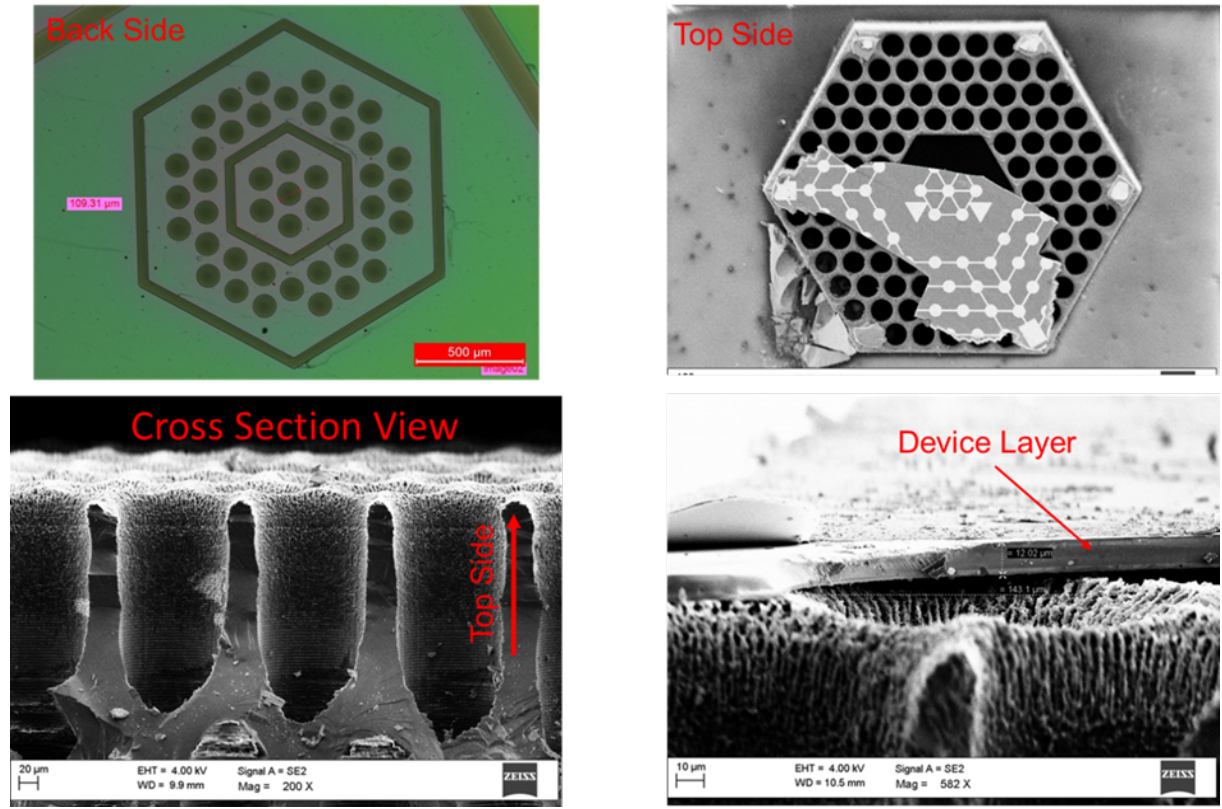

**Figure 3.8:** SEM micrograph of AlN-based PMUTs showing undercut after DRIE

### 3.4.2 Silicon Grass Formation

This problem of Si grass formation was observed in a few runs. This may be caused by incomplete etching of $SiO_2$ and the presence of polymer particles on the mask window created before the DRIE etching step. This problem was resolved by increasing the etching duration in the RIE step. Cross-sectional SEM of the device shows no grass formation.

## 3.5 Final Fabricated Device

After obtaining the right stack of material with good quality AlN as piezoelectric film and optimizing AlN etching, we decided to fabricate PMUT devices using AlN on SOI stack. Optical image of a released array of 6×6 PMUTs fabricated using the above mention method is shown in fig 3.9, It was observed that some diaphragms became curved/buckled. The reason for this can be residual stress in device stack and/or process variation during backside etching in the DRIE process [23].

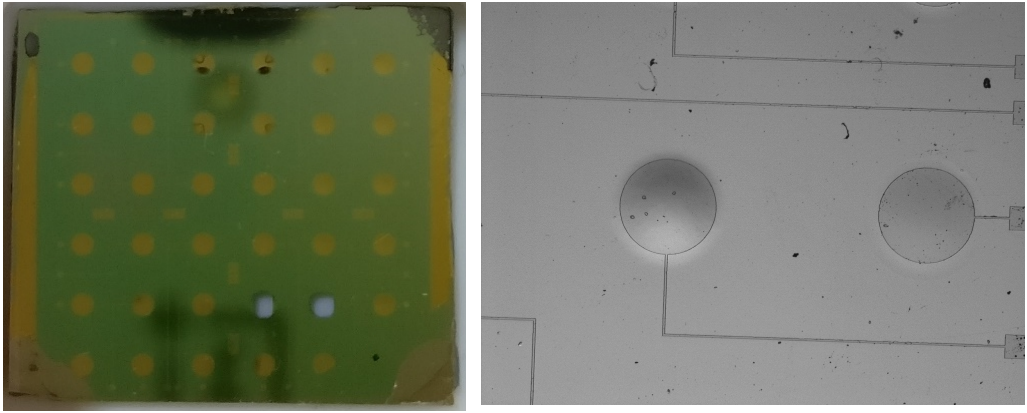

**Figure 3.9:** Optical image of a released array of 6×6 PMUTs

### 3.5.1 Optical Profilometry

To measure the static center deflection of diaphragm, Optical Profilometry was done to map 3D Profile of single PMUT device (see fig 3.10). From this, we measured the center deflection of the diaphragm and which came out to be in the range of 1 to 5μm. This variation in static deflection can cause variation in resonance frequency within the same array with the same diameter, which is not desirable and precautions should be taken during the future fabrication process to minimize this variation.

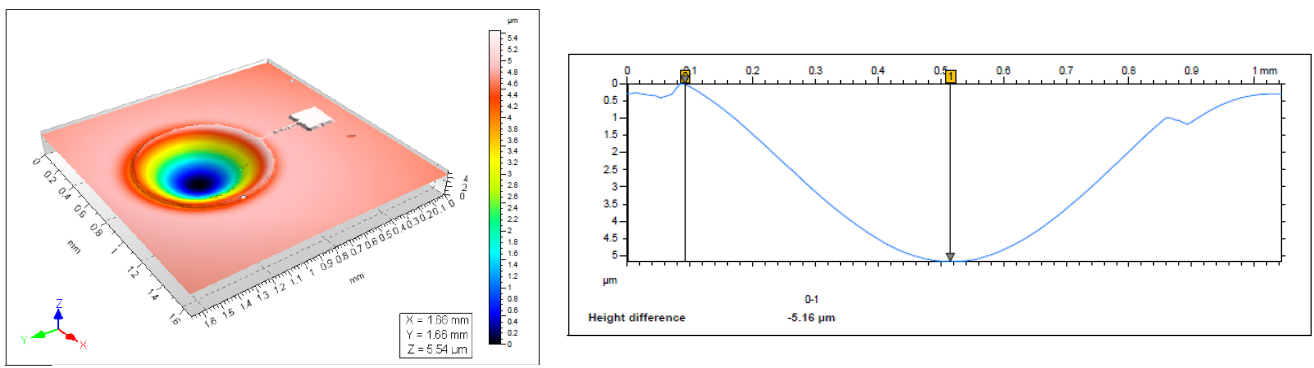

**Figure 3.10:** Optical Profilometry 3D Profile map of single PMUT device

## 3.5.2 Vibrational Characterization

The vibrational behavior of PMUTs captured using Micro System Analyser 500 (MSA500 by Polytec Inc) tool. Laser Doppler Vibrometry (LDV) was done to study the natural frequency of PMUT, their deflection sensitivity and the quality factors of the first resonance.

The wide frequency response of diaphragm type PMUTs shows the first resonance peak at 67.81kHz (Fig. 3.11).

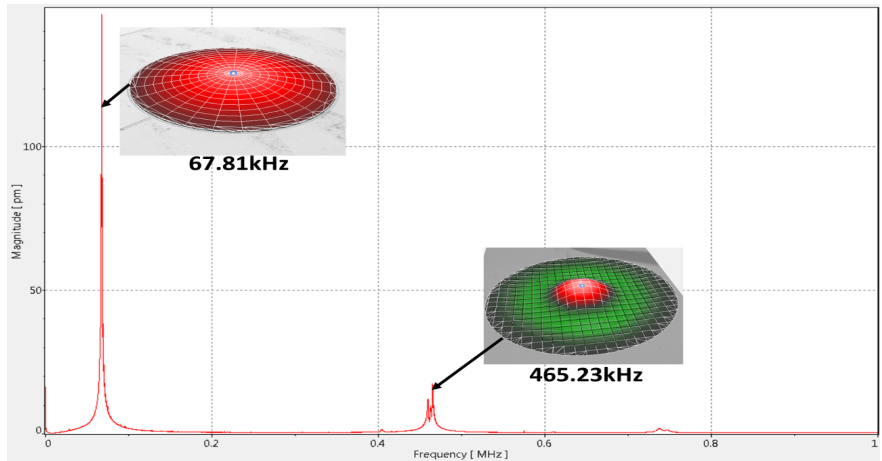

**Figure 3.11:** Wide Frequency response of the plate type PMUTs

Fig. 3.12 shows diaphragm deflection of 1mm diameter PMUT (10μm Device Layer) with varying AC voltages. From the graph, we can see that Center deflection Vs. Voltage is almost linear up to 10 V at least. At 1V, deflection is 118nm and at 10V, deflection is 1200nm. It can also be seen from the peak hold graph that, at higher voltages, frequency shifts to a higher value. This causes due to diaphragm becoming stiff at higher deflection values, which in turn exerts tensile stress on the diaphragm. Thus, causing resonance frequency to shift towards higher values.

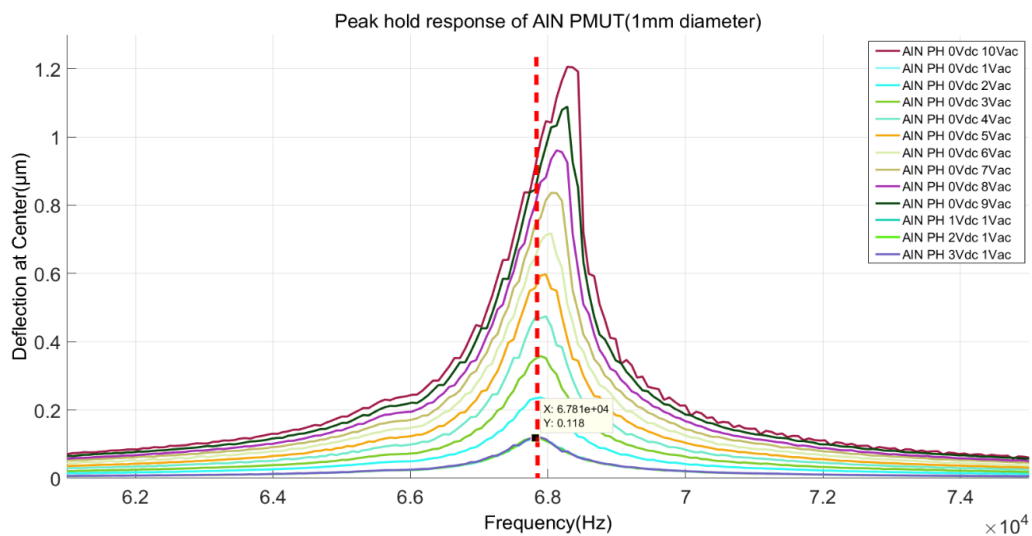

**Figure 3.12:** PMUT frequency response measured at various drive amplitudes.

# CHAPTER 4
# CONCLUSION

Complete development of an AlN-based PiezoMEMS device requires working on multiple fronts from material development to electronic integration. In this work, we have taken PMUTs as a prototype device to show proof of concept of working AlN-based PiezoMEMS device.

During this, we have explored all aspects of the development of a typical PMUTs starting from the development of thin films of the chosen piezoelectric material AlN, fabrication and characterization of PMUTs. Each of these parts presents challenges that belong to a different domain of engineering and can be dealt with at various depths. Here, we present our findings and shortcomings in each part one-by-one followed by our perspectives on the future possibilities in extending this work.

## 4.1 AlN Material development

We have successfully developed the stack of thin-film material with AlN an active piezoelectric layer for fabrication of a PMUT Array. Various challenges faced during the sputtering of AlN thin films and their solutions are discussed in details in the chapter. Multiple types of material failure such as AlN buckling, the stress in the film were resolved through rigorous process optimizations.

Following are the key conclusions drawn from our work on material development:

- Thicker piezoelectric AlN with thickness ~2μm was successfully deposited on SOI stack without buckling or cracking of film.

- By just varying the mode of power delivery to sputtering target from continuous DC to pulsed DC to RF, we can tune residual stress from -1.2Gpa (Compressive) to +250MPa (Tensile).

- During optimization of bottom electrode stack, we got a very good quality of platinum with minimum FWHM of Pt (111) =3°, which is comparable to commercially available platinized Silicon wafer with an added advantage in our case is it is deposited on SOI stack instead of Si. Which is crucial to make most of MEMS devices. This Pt optimization saved a lot of time and money, which may have been gone in purchasing platinized Silicon for PiezoMEMS device fabrication process.

- After optimizing AlN crystal quality, we achieved AlN (002) FWHM=4.5°. Which is very good when we see the dependence of piezoelectric coefficient on FWHM [12,13]. Below 5°, AlN piezoelectric coefficient value is ~99% of bulk single crystal value.

## 4.2 PMUTs array fabrication and characterization

After successfully depositing AlN based device stack, we fabricated and characterized plate type PMUTs with a diameter of 1000μm and device layer with thickness 10μm (p-D1000).

Following are the key conclusions drawn from our work on PMUT array fabrication and characterization:

- Wet etching AlN by 10% KOH at 90°C showed undercut i.e. lateral etching. This can be controlled by lowering the reaction temperature to 50°C.

- AlN etch rate by 10% KOH at 90°C along thickness was observed to be ~2μm/min and along lateral direction was ~4.2μm/min.

- AlN etch rate by 10% KOH at 50°C along thickness was observed to be ~300nm/min and along lateral direction was ~70nm/min.

- AlN etching by RIE recipe showed etch rate ~120nm/min.

- The selectivity of AlN: Pt was ~9:1 and AlN: PR(AZnOLF2020) was 0.5:1

- PMUT array (p-D1000) with 1000μm diameter was fabricated and characterized successfully. The resonance frequency of the first mode was measured to be 67.81kHz with quality factor ~90. Deflection sensitivity was calculated to be 118nm/V.

## 4.3 Future Work

- As we achieved AlN (002) FWHM ~5°, the next logical step will be to measure the piezoelectric coefficient with the help of the PNDS technique. We can also verify the $d_{33}$ and $d_{31}$ piezoelectric coefficient by fabricating cantilever structure.

- To improve Pt crystal quality and prevent diffusion of Pt into silicon, we are using 30 nm $TiO_2$ thin film deposited by ALD technique. But adhesion between oxide and metal is not so strong and this can cause delamination of the stack at $TiO_2$-Pt interface. So further efforts should be given to improve adhesion between $TiO_2$-Pt interface layers. One way to do this can be the introduction of 10nm thick sputtered Ti in-between $TiO_2$ and Pt layers.

- RIE recipe for AlN etching showed AlN spikes in SEM micrograph. This can be removed by wet etching. Further effort should be taken to standardize AlN etching recipe.

- Cantilever structure can also be used to calculate exact residual stress in device stack by measuring tip deflection and radius of curvature due to stress gradient in the multilayer stack.

## ACKNOWLEDGMENTS

The authors would like to thank Prof. Srinivasan Raghavan for his help and support. Authors also acknowledge Mr. Sandeep and Mr. Shashwat for providing us substrate material for platinum deposition. Authors acknowledge research funding from DST and MCIT, India.